\documentclass[twocolumn]{article}

\usepackage{algorithm}
\usepackage{amsthm}
\usepackage{amsfonts}
\usepackage{amssymb}
\usepackage{algorithm}
\usepackage{algorithmic}
\usepackage{amsmath}
\usepackage{graphicx}
\usepackage{setspace}
\usepackage{bm}
\usepackage{enumerate}
\usepackage{indentfirst}
\usepackage[colorlinks=false]{hyperref}
\usepackage{caption}
\usepackage{diagbox}
\usepackage{subfig}
\usepackage[multidot]{grffile}
\usepackage{booktabs}
\usepackage{palatino}
\usepackage{multirow}
\usepackage{balance}
\usepackage{titlesec}
\usepackage{parskip}
\usepackage{amsbsy}
\usepackage{mathrsfs}
\usepackage{authblk}
\DeclareMathAlphabet{\mathsfsl}{OT1}{cmss}{m}{sl}
\graphicspath{{figures/}{figure/}{pictures/}{picture/}{pic/}{pics/}{image/}{images/}}
\title{Key technologies and application for radar and smart video fusion in perimeter intrusion alarm system}
\author[1,*]{Shujun Fu}
\author[1]{Shenghai Liao}
\author[2]{Jingjing Gao}
\author[2]{Shijing Song}
\author[2]{Zhonghua Man}
\affil[1]{School of Mathematics, Shandong University, Jinan 250100, China}
\affil[2]{Shandong Feitian photoelectric technology limited company, Tengzhou 277500, China}
\affil[*]{Correspondence: shujunfu@163.com}
\usepackage{geometry}
\usepackage{indentfirst}
\date{August 26, 2023}

\begin{document}
\maketitle
\begin{abstract}
With the continuous development of modern science and technology, radar detection, video surveillance and perimeter alarm system are more and more widely used in the field of social security. This paper introduces video surveillance and perimeter alarm in detail, mathematical modeling and key technologies, analyzes their fusion and application status, and puts forward suggestions combined with the development trend of intelligent security system in the future.
\end{abstract}


\section{Introduction}

Radar comes from the abbreviation of radio detection and ranging, meaning "radio detection and ranging", that is, to find targets by radio methods and determine their spatial positions. Radar is an electronic device that uses electromagnetic wave to detect the target. The radar emits electromagnetic wave to irradiate the target and receive its echo, so as to obtain the distance from the target to the electromagnetic wave transmitting point, the radial velocity, the orientation, the height and other information.

Through the installation of cameras and other equipment, by real-time video acquisition, transmission, storage and playback and other technical means, video surveillance achieves the security monitoring of a designated area and post-investigation and forensics and other functions.

Perimeter intrusion alarm system (PIAS) is a special system for perimeter security, mainly divided into infrared alarm, laser alarm, electronic fence and so on \cite{lohani2022perimeter}. Through the installation of sensors, infrared and other equipment, by real-time monitoring and alarm, perimeter alarm achieves the perimeter security and post-investigation and other functions of a designated perimeter.

Radar detection can be monitored in harsh weather and environmental conditions, is able to detect fast-moving objects;  it has a wide range of monitoring, high reliability. But the resolution for fixed targets is not as good as video surveillance. In perimeter alarm system video surveillance can obtain detailed target information, such as appearance, shape, etc. It can be used in combination with other security systems, such as access control systems. However, the limited monitoring capacity in harsh weather and environmental conditions, and the requirement of a lot of manual monitoring hinder its development.

In the perimeter intrusion alarm system with radar and smart video fusion, long transmission distance, strong scalability and remote monitoring can be achieved with relatively low false and missed alarms\cite{alessandretti2007vehicle,chang2020spatial}.

Video surveillance and perimeter alarm systems have a wide range of application prospects in various fields, the following are several typical areas\cite{qiang2020research}.

Urban safety and public safety. With the acceleration of urbanization, more and more attention has been paid to urban security. Video surveillance and perimeter alarm system can provide a full range of security monitoring and early warning services for the city, effectively prevent and combat criminal behavior. At the same time, in public places such as shopping malls, stations, airports, etc., video surveillance and perimeter alarm systems can provide security and maintain social order and public safety.

Transportation field. Video surveillance and perimeter alarm systems have a wide range of applications in the field of transportation, such as highways, parking lots, railway stations and so on. The system can monitor traffic flow and vehicle driving in real time, and provide data support for traffic management and scheduling. At the same time, the video surveillance and perimeter alarm system can ensure the safety of traffic facilities and vehicles and reduce the accident rate.

Environmental protection and energy management. Video surveillance and perimeter alarm systems can monitor and record environmental changes and energy use, providing data support for environmental protection and energy management. For example, through the video surveillance system, the situation of forests, grasslands and other nature reserves can be monitored in real time, and environmental damage can be detected and dealt with in a timely manner. At the same time, through the perimeter alarm system, the security of energy facilities can be monitored to ensure the stability and security of energy supply.

Healthcare and smart home. In the field of health care, video surveillance and perimeter alarm systems can provide security and telemedicine services for hospitals, nursing homes, etc. Through real-time monitoring of patients' condition and vital signs, changes in condition can be detected and dealt with in time to improve medical quality and efficiency. At the same time, in the field of smart home, video surveillance and perimeter alarm systems can achieve home intelligent management and remote control, improving home security and convenience of life.

In short, video surveillance and perimeter alarm system has a wide range of application prospects in various fields, with the continuous development and application of new technology, its application field will continue to expand and deepen. This will bring more innovation and development opportunities for social development.

\section{Perimeter intrusion alarm system}

\begin{figure*}
\centering
\newcommand{\fs}{0.99}
\subfloat{
\label{pp}
\includegraphics[width=\fs\linewidth]{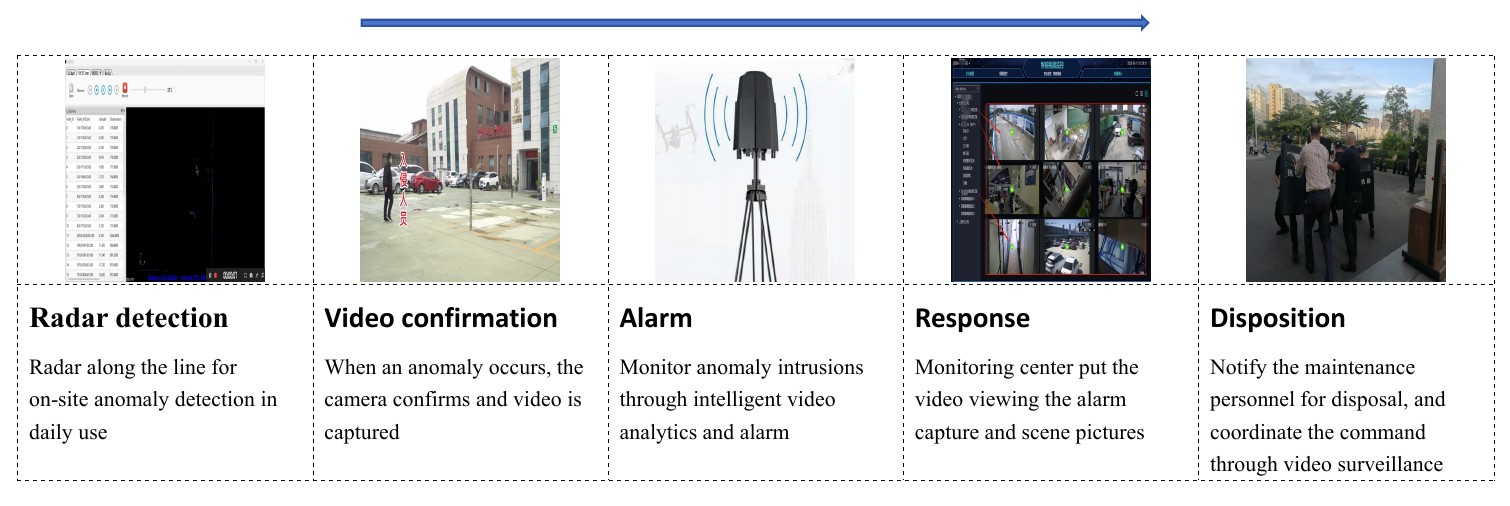}
}
\caption{Main steps of a perimeter intrusion alarm system with fusion of radar and smart video, including radar detection, video confirmation, alarm, response and disposion.}\label{pias}
\end{figure*}

The typical pipeline of a perimeter intrusion alarm system involves several stages (see figure \ref{pp}). Here is a detailed overview of the process \cite{qiang2020research}. It is important to note that the specific details of the pipeline may vary depending on the specific PIAS implementation and the requirements of the protected area.

1. Sensor deployment: the first step is to deploy sensors along the perimeter that needs to be protected. These sensors can include technologies such as infrared, microwave, acoustic, or video-based sensors. They are strategically placed to detect any unauthorized entry or intrusion attempts.

2. Sensor data acquisition: once the sensors are deployed, they continuously monitor the perimeter and collect data. This data can include information such as motion, heat signatures, sound, or video footage, depending on the type of sensors used.

3. Data processing: the collected sensor data is then processed to extract meaningful information. This can involve filtering out noise, analyzing patterns, and identifying potential intrusion events. Advanced algorithms and machine learning techniques can be employed to improve the accuracy of intrusion detection.

4. Event detection: in this stage, the processed data is analyzed to detect potential intrusion events. This can involve comparing the collected data against predefined thresholds or using anomaly detection algorithms to identify abnormal behavior. If an intrusion event is detected, an alert is generated.

5. Alert generation: when an intrusion event is detected, an alert is generated to notify security personnel or a central monitoring station. The alert can be in the form of a visual or audible alarm, email notification, or integration with a security management system.

6. Intrusion confirmation: once an alert is generated, security personnel or the monitoring station verifies the intrusion event. This can be done through manual verification, remote video surveillance, or by deploying response teams to the location.

7. Response and mitigation: if an intrusion is confirmed, appropriate response measures are taken to mitigate the threat. This can include dispatching security personnel, activating security barriers or locks, or initiating an emergency response protocol.

8. Logging and reporting: throughout the entire process, detailed logs are maintained to record all relevant events, including sensor activations, alerts, and response actions. These logs can be used for forensic analysis, compliance reporting, or system performance evaluation.

In a perimeter alarm system of with fusion of radar and smart video, video analysis and object detection are relatively difficult. In the following sections, we mainly discuss background subtraction and video analysis.

\section{Modeling of PIAS for video surveillance}

The task of perimeter intrusion alarm system (PIAS) can be defined as a monitoring system that identifies the presence
of humans or devices in a pre-defined field-of-view. Modeling of PIAS defines mathematically a PIAS with clarity and formalization \cite{lohani2022perimeter}.

A video $\mathcal{V}$ acquired for $T$ frames during the interval $\mathcal{T}=[1, T]$ is defined as:
\begin{equation}\label{}
  \mathcal{V}=\left\{I_t\in \mathbb{R}^{H\times W\times D}, t\in \mathcal{T}\right\}
\end{equation}
where $I_t$ denotes the frame at the time instant $t$, with height $H$, width $W$ and number of channels $D$.

Including all of the frames where it is present, an object $oi$ in the video is defined as:
\begin{equation}\label{}
  o_i=\left(\{b_{i,t}\}, c_i\in \mathcal{C}, t\in \mathcal{T}\right),\ b_{i,t}=\{g_{i,t}, w_{i,t}, h_{i,t}\}
\end{equation}
where $c_i$ is the class of the object from the set of object classes $\mathcal{C}$, and $b_{i,t}$ is its bounding box at time instant $t$; $g_{i,t}, w_{i,t}$ and $h_{i,t}$ are the center, width and height of the bounding box,
respectively.

An object causes an intrusion event if it belongs to a non-authorized class and is moving in a protected area during a prohibited time interval \cite{lohani2022perimeter}. The intrusion event caused by an object $o_i$ is defined as:
\begin{align}\label{}
  \mathcal{IE}(o_i)=&\left\{I_t, c_i\in \mathcal{C}_{na}, t\in \mathcal{T}_{na}, \|grad\ g_{i,t}\|>0, \right. \notag\\
  &\left. g_{i,t}\in \mathcal{S}_{na}, t\in \mathcal{T}\right\}
\end{align}
where $\|grad\ g_{i,t}\|$ is the gradient of object $o_i$. $\mathcal{S}_{na}, \mathcal{T}_{na}$ and $\mathcal{C}_{na}$ define the boundary to protect, time interval and classes for non-authorized (na) intruders.

Given a video $\mathcal{V}$ and parameters ($\mathcal{S}_{na}, \mathcal{T}_{na}$, $\mathcal{C}_{na}$), the
prediction of a PIAS is defined as:
\begin{equation}\label{}
  \mathcal{P}(\mathcal{V}, \mathcal{S}_{na}, \mathcal{T}_{na}, \mathcal{C}_{na})=\left\{\hat{p}_t\in \{0, 1\}, t\in \mathcal{T}_{na}\right\}
\end{equation}
where $\hat{p}_t$ is a binary prediction for each frame $t$ of the video $\mathcal{V}$ for time $\mathcal{T}_{na}$.

Given the precise definitions of the intrusion event and the prediction of PIAS, one can evaluate the performance of a PIDS by comparing the PIDS output with ground truth annotations.

\section{Key technologies for PIAS}

Video surveillance and perimeter alarm systems involve a number of related technologies, the following are a few important technologies that we have studied recently, specially in the context of background subtraction \cite{piccardi2004background,kalsotra2022background}.

\subsection{Image and video denoising}

Denoising is an important step in background subtraction to reduce noise and improve the accuracy of foreground object detection. Here are some denoising techniques commonly used in PIAS \cite{buades2005review,xie2016weighted}.

Gaussian smoothing: gaussian smoothing, also known as Gaussian blur, is a widely used technique for reducing noise in images. It applies a Gaussian filter to the input frames, which effectively blurs the image and reduces high-frequency noise. This can help in obtaining a smoother background model for background subtraction.

Median filtering: median filtering is a non-linear filtering technique that replaces each pixel value with the median value of its neighboring pixels. It is particularly effective in removing salt-and-pepper noise, which appears as isolated white and black pixels. Median filtering can help in preserving edges and details while reducing noise in the background model.

Bilateral filtering: bilateral filtering is a non-linear filtering technique that preserves edges while reducing noise. It considers both the spatial distance and intensity difference between pixels when applying the filter. Bilateral filtering can effectively smooth the background while preserving the sharpness of foreground object boundaries.

Non-local means denoising \cite{buades2008nonlocal}: non-local means denoising is a technique that exploits the redundancy in image patches to reduce noise. It compares similar patches in the image and averages their pixel values to estimate the denoised value for each pixel. Non-local means denoising can effectively remove noise while preserving fine details in the background model.

Total variation denoising \cite{rudin1992nonlinear}: total variation denoising is a technique that minimizes the total variation of an image while preserving important features. It reduces noise by promoting piecewise smoothness in the image.

Wavelet denoising: wavelet denoising is a technique that utilizes the wavelet transform to decompose an image into different frequency bands. By thresholding the coefficients in the high-frequency bands, wavelet denoising can effectively reduce noise while preserving important image features. This technique is particularly useful for denoising images with both high-frequency noise and low-frequency background information.

Non-Local sparse representation: non-local sparse representation methods exploit the sparsity of the foreground objects in the image. By representing the image as a linear combination of a few atoms from a learned dictionary, these methods can effectively separate the foreground objects from the background.

Non-local low-rank matrix completion: it is a denoising technique that combines non-local means denoising with low-rank matrix completion. It exploits the low-rank structure of the background and the non-local similarity of image patches to recover a clean background model. It can effectively remove noise while preserving the global structure of the background.

Deep learning-based denoising \cite{xie2012image}: deep learning approaches, such as convolutional neural networks (CNNs), have shown promising results in denoising tasks. These models are trained on large datasets to learn the mapping between noisy and clean images. Deep learning-based denoising methods can effectively remove noise while preserving fine details in the background model.

Temporal denoising: in background subtraction, temporal denoising techniques exploit the temporal coherence of consecutive frames to reduce noise. Methods like temporal averaging or temporal median filtering can be applied to smooth out temporal variations and improve the accuracy of background subtraction.

Non-Local low-rank tensor decomposition: these methods extend low-rank techniques to higher-dimensional data, such as video sequences. These methods exploit the low-rank structure of the background tensor and the non-local similarity of video patches to denoise the background and separate it from the foreground objects.

These denoising techniques can be applied to the input frames before or during the background modeling process in background subtraction. It's important to note that the choice of denoising technique depends on the specific characteristics of the noise, the desired level of noise reduction, and the trade-off between noise removal and preservation of important image features. Experimentation and evaluation on the specific dataset are crucial to determine the most suitable denoising technique for background subtraction and intrusion event detection.

\subsection{Enhancement of low-light image and video}

Enhancing low-light conditions for background subtraction can be challenging due to the limited amount of available light. However, there are several techniques that can help improve the visibility of the background and enhance the accuracy of background subtraction in low-light scenarios \cite{li2021low,ai2020extreme,lamba2021restoring}.

Histogram equalization: histogram equalization is a simple yet effective technique for enhancing the contrast of an image. It redistributes the pixel intensities to cover the entire dynamic range, which can help reveal details in low-light areas. Applying histogram equalization to the input frames can enhance the visibility of the background and improve background subtraction results.

Adaptive histogram equalization: adaptive histogram equalization is an extension of histogram equalization that adapts the equalization process to local image regions. By dividing the image into smaller regions and applying histogram equalization independently to each region, adaptive histogram equalization can enhance the visibility of both bright and dark regions in low-light images.

Retinex-based methods: retinex-based methods aim to restore the illumination and reflectance components of an image. These methods separate the image into these two components and enhance the reflectance component, which represents the underlying scene information. By enhancing the reflectance component, retinex-based methods can improve the visibility of the background in low-light conditions.

Multi-exposure fusion: multi-exposure fusion techniques combine multiple images captured with different exposure settings to create a single image with enhanced dynamic range. By fusing the information from multiple exposures, these techniques can reveal details in both dark and bright regions of the image. Multi-exposure fusion can be particularly useful for enhancing the visibility of the background in low-light conditions.

Deep learning-based methods \cite{li2021low,ai2020extreme}: deep learning approaches have shown promising results in low-light image enhancement. These models are trained on large datasets to learn the mapping between low-light and well-exposed images. By applying deep learning-based methods, the visibility of the background can be improved in low-light conditions, leading to better background subtraction results.

Noise reduction: in low-light conditions, noise can be more prominent. Applying denoising techniques, such as those mentioned earlier, can help reduce noise and improve the visibility of the background. By reducing noise, the accuracy of background subtraction can be enhanced.

Illumination compensation: illumination compensation techniques aim to equalize the illumination across the image. These methods estimate the illumination map and adjust the pixel values accordingly.

Low-light image fusion: low-light image fusion combines multiple low-light images of the same scene to create a single enhanced image. This technique leverages the information from multiple images to improve the visibility of the background. Fusion methods can be based on averaging, weighted averaging, or more advanced algorithms like Laplacian pyramid fusion.

Image enhancement filters: various image enhancement filters, such as gamma correction, contrast stretching, and adaptive filtering, can be applied to enhance the visibility of the background in low-light conditions. These filters adjust the pixel values to improve contrast and reveal details in dark areas.

Infrared imaging: infrared imaging can be used as an alternative to visible light imaging in low-light conditions. Infrared cameras capture thermal radiation emitted by objects, which can provide better visibility in low-light or no-light environments. By using infrared imaging, the background can be captured more clearly, leading to improved background subtraction results.

It's important to note that the choice of technique depends on the specific characteristics of the low-light conditions, the available equipment, and the desired trade-off between noise reduction, contrast enhancement, and visibility improvement. Evaluating the performance of different techniques on the specific dataset and considering the limitations of the hardware are crucial for selecting the most suitable approach for enhancing low-light conditions in background subtraction.

\subsection{low-rank methods}

There are several low-rank methods that can be used for background subtraction in computer vision applications\cite{javed2016spatiotemporal,LIU201985}. Here are a few commonly used ones.

Principal component analysis (PCA): PCA is a widely used technique for dimensionality reduction. In the context of background subtraction, PCA can be applied to a set of training images to extract the principal components that capture the background information. The foreground objects can then be detected by subtracting the background model from the input frames.

Robust principal component analysis (RPCA) \cite{candes2009exact}: RPCA extends PCA by incorporating a low-rank and a sparse component. The low-rank component represents the background, while the sparse component represents the foreground objects. By decomposing the input frames into these two components, foreground objects can be effectively separated from the background.

Low-rank and sparse decomposition (LRSD): LRSD is a method that decomposes the input frames into a low-rank component representing the background and a sparse component representing the foreground objects. It leverages the fact that the background is typically low-rank, while the foreground objects introduce sparsity in the data.

Online robust subspace tracking (ORST): ORST is an online method that tracks the background subspace over time. It updates the background model incrementally, allowing it to adapt to gradual changes in the scene. ORST combines low-rank and sparse decomposition techniques to separate the background from the foreground objects.

Background subtraction via low-rank and structured sparse decomposition (BSSSD): BSSSD is a method that combines low-rank and structured sparse decomposition techniques. It exploits the structured sparsity of the foreground objects, such as their spatial or temporal coherence, to improve the accuracy of background subtraction.

Low-rank representation with local constraint (LRRLC): LRRLC is a method that combines low-rank representation with local constraints to improve background subtraction. It considers the local spatial and temporal information of the input frames to enhance the accuracy of separating the background from the foreground objects.

Low-rank tensor decomposition (LRTD): LRTD extends low-rank methods to handle multi-dimensional data, such as video sequences. It decomposes the input frames into a low-rank tensor representing the background and a sparse tensor representing the foreground objects. This allows for background subtraction in higher-dimensional data.

These methods provide different strategies for background subtraction, taking into account various factors such as adaptability to changes, robustness to outliers, local constraints, multi-dimensional data, online processing, structured sparsity and matrix decomposition. It is important to evaluate and choose the most suitable method based on the specific requirements and characteristics of your application.

\section{Main issues and prospects}

\subsection{Main issues}

Applying video surveillance to perimeter protection alarms can be a complex task, and there are several key issues and difficulties that need to be addressed. Here are some of main challenges \cite{chen2019distributed,sreenu2019intelligent,tsakanikas2018video}.

False alarms: one of the primary challenges in video surveillance for perimeter protection is the occurrence of false alarms. False alarms can be triggered by various factors such as environmental conditions (e.g., moving trees or shadows), animals, or even system malfunctions. Reducing false alarms is crucial to ensure the effectiveness and reliability of the system.

Variable lighting conditions: lighting conditions can vary significantly in outdoor environments, posing a challenge for video surveillance systems. Changes in lighting, such as shadows, glare, or low-light conditions, can affect the accuracy of object detection and tracking algorithms. Developing robust algorithms that can handle varying lighting conditions is essential.

Occlusions and clutter: perimeter areas can have various objects, structures, and vegetation that may obstruct the view or cause clutter in the video footage. Occlusions and clutter can make it difficult to accurately detect and track objects of interest, potentially leading to missed intrusions or false alarms. Advanced algorithms that can handle occlusions and clutter are needed.

Camera placement and coverage: determining the optimal placement and coverage of cameras is crucial for effective perimeter protection. It requires careful consideration of factors such as the layout of the area, potential blind spots, and the range and resolution of the cameras. Finding the right balance between camera placement and coverage is essential to maximize the system's effectiveness.

Scalability and cost: deploying video surveillance systems for perimeter protection over large areas can be challenging in terms of scalability and cost. It requires a significant number of cameras, storage capacity, and computational resources. Managing and maintaining such systems can be costly and resource-intensive.

Privacy concerns: video surveillance systems raise privacy concerns, as they capture and record visual information of individuals in public or private spaces. Balancing the need for security with privacy rights is a critical challenge. Compliance with privacy regulations and implementing privacy-enhancing measures, such as video anonymization or masking, is necessary to address these concerns.

Data storage and management: video surveillance systems generate a vast amount of data that needs to be stored and managed effectively. This includes considerations such as data retention policies, storage capacity, data backup, and retrieval. Efficient data management strategies, including compression techniques and cloud-based storage solutions, are essential to handle the large volume of video data.

Real-time processing: perimeter protection alarms require real-time processing and analysis of video streams to detect and respond to intrusions promptly. This places high demands on the computational capabilities of the system. Efficient algorithms and hardware acceleration techniques, such as GPUs or dedicated video analytics processors, are necessary to achieve real-time processing.

Integration with other security systems: perimeter protection alarms are often part of a larger security ecosystem that includes access control systems, intrusion detection systems, and alarm management platforms. Integrating video surveillance with these systems to provide a comprehensive security solution can be challenging. Interoperability, data sharing, and seamless integration are key considerations in achieving a unified security infrastructure.

System reliability and maintenance: video surveillance systems for perimeter protection require regular maintenance and monitoring to ensure their reliability. This includes tasks such as camera calibration, firmware updates, system diagnostics, and fault detection. Implementing proactive maintenance strategies and remote monitoring capabilities can help minimize system downtime and ensure continuous operation.

Compliance with regulations and standards: video surveillance systems must comply with various regulations and standards, such as data protection laws and industry-specific guidelines. Ensuring compliance can be challenging, especially when operating in multiple jurisdictions or industries with different requirements. Adhering to privacy regulations, data protection practices, and industry standards is essential to maintain the legal and ethical use of video surveillance systems.

Addressing these key issues and difficulties requires a multidisciplinary approach, involving expertise in computer vision, data management, privacy, security, and system integration. Researchers and practitioners continue to work on developing innovative solutions and best practices to overcome these challenges and improve the effectiveness and efficiency of video surveillance for perimeter protection alarms.

\subsection{Prospects and proposals}

According to the development trend of video surveillance and perimeter intrusion alarm system in the future, the following suggestions are put forward.

Improve the intelligent level of the system. Strengthen the application research and development of artificial intelligence technology in the field of video surveillance and perimeter alarm, improve the intelligence level of the system, and improve the identification accuracy rate and early warning timeliness. Two particular points worth making are multidimensional data acquisition and fusion, and object recognition and tracking based on deep learning.

Strengthen technological research, development and innovation. Strengthen technology research and development and innovation, and promote the technological upgrading and upgrading of video surveillance and perimeter alarm systems. At the same time, strengthen industry-university-research cooperation and train technical talents in related fields.

Improve system security and stability. Strengthen the security and stability of the system to prevent risks such as hacker attacks and data leaks. At the same time, strengthen the reliability and stability of the equipment, reduce the failure rate and false alarm rate.

Strengthen policy guidance and industry norms. Strengthen policy guidance and industry norms to promote the healthy development of video surveillance and perimeter alarm industries. At the same time, strengthen the formulation and implementation of relevant regulations to standardize industry standards and market.

\section{Conclusion}

Video surveillance and perimeter instrusion alarm system is an important part of the security field, which is of great significance to ensure social security. In the future, with the continuous development of artificial intelligence and Internet of Things technology, video surveillance and perimeter alarm systems will be more intelligent, diversified and networked. At the same time, strengthening technology research and development and innovation, improving the security and stability of the system, and strengthening policy guidance and industry norms are the key to promoting the healthy development of the video surveillance and perimeter alarm industry.

\section*{Acknowledgment}

The research has been supported in part by the talent project of Shandong Province Development and Reform Commission (Radar-vison integrated intelligent intrusion prevention technology research and industrial applications, 2022-23),
the National Natural Science Foundation of China (12071263) and the Natural Science Foundation of Shandong Province of China (ZR2019MF045).

\setlength{\parskip}{\baselineskip}
\bibliographystyle{ieeetr}
\bibliography{mylib}

\end{document}